\documentclass[conference,a4paper]{IEEEtran}
\IEEEoverridecommandlockouts
\usepackage{stfloats,color}
\usepackage{amsfonts}
\usepackage[cmex10]{amsmath}
\usepackage{graphicx}
\usepackage{setspace}
\usepackage{cite}
\usepackage{array}
\usepackage{algorithmic}
\usepackage{mdwmath}
\usepackage{mdwtab}
\usepackage[center]{caption}
\usepackage[font=footnotesize]{subfig}
\usepackage{fixltx2e}
\usepackage{url}
\usepackage{times}
\usepackage{epsfig}
\usepackage{latexsym}
\usepackage{epstopdf}
\usepackage{verbatim}
\usepackage{units}
\usepackage{amsthm}
\usepackage{mdwlist}

\begin{document}
\title{Interactive Multiple Model Estimation of Doubly-Selective Channels for OFDM systems}

\author{\IEEEauthorblockN{Mahmoud Ashour$^\dag$ and Amr El-Keyi$^{\ddag}$}

\IEEEauthorblockA{$^\dag$ Electrical Engineering Dept., Pennsylvania State University, PA, USA.\\
$^\ddag$ Department of Systems and Computer Engineering, Carleton University, Ottawa, Canada.
}
}

\maketitle

\begin{abstract}
In this paper, we propose an algorithm for channel estimation, acquisition and tracking, for orthogonal frequency division multiplexing (OFDM) systems. The proposed algorithm is suitable for vehicular communications that encounter very high mobility. A preamble sequence is used to derive an initial estimate of the channel using least squares (LS). The temporal variation of the channel within one OFDM symbol is approximated by two complex exponential basis expansion models (CE-BEM). One of the Fourier-based BEMs is intended to capture the low frequencies in the channel (slow variations corresponding to low Doppler), while the other is destined to capture high frequencies (fast variations corresponding to high Doppler). Kalman filtering is employed to track the BEM coefficients iteratively on an OFDM symbol-by-symbol basis. An interactive multiple model (IMM) estimator is implemented to dynamically mix the estimates obtained by the two Kalman filters, each of which matched to one of the BEMs. Extensive numerical simulations are conducted to signify the gain obtained by the proposed combining technique.    
\end{abstract}

\IEEEpeerreviewmaketitle

%\vspace{-7.5pt}

\section{Introduction}
\IEEEPARstart{O} \small{r}thogonal frequency division multiplexing (OFDM) is famous for its robustness against frequency-selective wireless channels. Spectrally shaped channels are transformed into a set of parallel flat-fading subchannels to enable high data rate transmission. OFDM is widely used in new emerging standards for mobile communications, e.g., long-term evolution (LTE) and WiMax. However, the design of OFDM systems in the presence of high mobility is challenging. The relative motion between the transmitter and receiver induces temporal variations in the channel which destroys orthogonality among the subcarriers. Thus, OFDM receivers require accurate channel information for reliable detection of the transmitted signals.      

Accurate channel estimation in the presence of high mobility conditions has received a lot of attention \cite{RLS,Nafouri}. A pilot-aided technique is developed in \cite{RLS}, where the received signal samples containing pilot tones in the frequency domain are used to estimate and track the channel impulse response (CIR) coefficients via a recursive least-squares (RLS) algorithm. In \cite{Nafouri}, the authors propose a pilot-aided technique that exploits the channel correlations in time and frequency domains. Most of the computations are performed offline to reduce the complexity of the algorithm. 

Another widely investigated trend is the use of parametric channel models. The temporal variation of the channel coefficients within one OFDM symbol is approximated by a basis expansion model (BEM). The time-varying channel taps are represented as a weighted sum of time-varying basis functions. Among the candidate basis functions are the complex exponential (Fourier) functions \cite{Giannakis,subblock}, polynomials \cite{QR}, and discrete Karhuen-Loeve functions that correspond to the most significant eigenvectors of the channel correlation matrix \cite{KL_BEM}.

In this paper, we propose an algorithm for channel estimation, acquisition and tracking, in the presence of high mobility. The algorithm uses a preamble sequence to derive an initial estimate of the channel coefficients using least squares (LS). Then, the temporal variation of the channel coefficients are modelled by two complex exponential BEMs. One of the Fourier-based BEMs is intended to capture the low frequencies in the channel (slow variations corresponding to low Doppler), while the other aims at capturing high frequencies (fast variations corresponding to high Doppler). Kalman filtering proposed in \cite{MAP} is then used to track the BEM coefficients under each individual BEM. An interactive multiple model (IMM) estimator is implemented to dynamically combine the estimates obtained by the two Kalman filters. It is shown through numerical simulations that the proposed combining technique outperforms the use of only one Kalman filter matched to a BEM containing a concatenation of the basis vectors available in both of the used BEMs.

\subsubsection*{Notation}
We denote scalars by lower-case letters (e.g. $x$), vectors by lower-case boldface letters (e.g. $\mathbf{x}$) and matrices by upper-case boldface letters (e.g. $\mathbf{X}$). A hat over a variable refers to its estimate (e.g. $\hat{x}$ is an estimate of $x$). Superscripts $T$ and $H$ denote
transpose and hermitian transpose, respectively. We reserve $\mathbb{E}$ for the statistical
expectation operator. The $N \times N$ identity matrix is denoted by $\mathbf{I}_{N}$. The $(k,m)$th entry of the matrix $\mathbf{X}$ is denoted by $[\mathbf{X}]_{k,m}$. A diagonal matrix with $\mathbf{x}$ on its main diagonal is denoted by $\text{diag}\{ \mathbf{x} \}$. $\text{blkdiag}\{ \mathbf{X},\mathbf{Y} \}$ is a block diagonal matrix with $\mathbf{X}$ and $\mathbf{Y}$ on its main diagonal.

\section{System Model}
In this section, we present a matrix-vector model for a discrete-time baseband-equivalent OFDM system. We consider a system with N subcarriers. Let $\mathcal{X}_{n}$ denote the frequency-domain transmitted symbols of the $n$th OFDM symbol. The corresponding time-domain samples, $\mathbf{x}_{n}$, are obtained through an N-point inverse discrete Fourier transformation (IDFT) of $\mathcal{X}_{n}$. Thus, $\mathbf{x}_{n}=\mathbf{F}^{H} \mathcal{X}_{n}$. $\mathbf{F}$ denotes the N-point DFT matrix.      
A cyclic prefix (CP) of length $N_g$ is augmented to the time-domain transmitted samples to avoid inter-symbol interference (ISI) between consecutive OFDM symbols. Therefore, the OFDM symbol duration is given by $T=N_{s}T_{s}$, where $T_{s}$ denotes the sampling interval of the system and $N_{s}=N+N_{g}$. After transmission over a multipath fading channel with $L < N_g$ taps, the received time-domain signal (after CP removal) corresponding to the $n$th OFDM symbol is written following the model presented in \cite{Nafouri} as
\begin{align}\label{sysmodel}
\mathbf{y}_{n}=\mathbf{H}_{n} \mathbf{x}_{n} + \mathbf{w}_{n}
\end{align}
where $\mathbf{H}_{n}$ is the $N \times N$ time-domain channel matrix with the following structure
\begin{align} \label{ch_mtx} 
\mathbf{H}_{n}\!\!=\!\!
\begin{bmatrix}
  h_{0,n}(0)     & 0             & \ldots   & h_{1,n}(0)\\
  h_{1,n}(1)     & h_{0,n}(1)    & \ldots   & h_{2,n}(1) \\
  \vdots         &\ddots         & \ddots   & \vdots      \\
  h_{L-1,n}(L-1) & h_{L-2,n}(L-1)& \ldots   & 0            \\
  0              & h_{L-1,n}(L)  & \ldots   & 0             \\
  \vdots         & 0             & \ddots   & \vdots         \\
  0              & \vdots        & \ldots   & 0               \\
  0              & 0             & \ldots   & h_{0,n}(N-1)  
\end{bmatrix}\quad \quad \quad \quad
\end{align}
and $h_{l,n}(q)$ is the complex channel gain of the $l$th tap in the $n$th OFDM symbol at time instant $q$, and $\mathbf{w}_{n}$ is the time-domain additive zero-mean complex Gaussian noise with covariance matrix $\sigma_{\mathbf{w}}^{2}\mathbf{I}_{N}$ added to the $n$th received OFDM symbol. Applying DFT to (\ref{sysmodel}), we obtain the frequency domain received symbols
\begin{align}
\mathcal{Y}_{n}=\mathbf{G}_{n}\mathcal{X}_{n}+\mathcal{W}_{n}
\end{align} 
where $\mathbf{G}_{n}=\mathbf{F}\mathbf{H}_{n}\mathbf{F}^{H}$ is the frequency domain channel matrix at the $n$th OFDM symbol, and $\mathcal{W}_{n}=\mathbf{F}\mathbf{w}_{n}$ is the frequency domain noise vector added to the $n$th OFDM symbol. 

If the channel is quasi-static, i.e., the channel coefficients are constant within the same OFDM symbol, then the channel matrix $\mathbf{H}_{n}$ becomes circulant. Hence, it can be diagonalized through applying DFT and IDFT operations yielding a diagonal matrix $\mathbf{G}_{n}$. Thus, the subcarriers keep their orthogonality and no inter-carrier interference (ICI) is introduced. However, in very high mobility conditions, the relative motion between the transmitter and the receiver induces temporal variations in the channel. Therefore, the matrix $\mathbf{H}_{n}$ can no longer be considered circulant. This introduces off-diagonal elements to the corresponding matrix $\mathbf{G}_{n}$ causing ICI. When the channel is fast time varying, ICI becomes significant and severely degrades the system performance. This stimulates the necessity of estimating the channel coefficients at every sampling time instant. Next, we introduce the BEM that helps us do that job.

\section{Basis Expansion Model}
In this section, we introduce the approach of using a BEM to model the temporal variation of the channel coefficients within one OFDM symbol. The channel taps are considered wide sense stationary (WSS) zero-mean complex Guassian processes of variances $\{\sigma_{h_{l}}^{2}\}_{l=0}^{L-1}$. The N-dimensional vector of the $l$th channel tap at the $n$th OFDM symbol is defined as
\begin{align}
\mathbf{h}_{l,n}= [h_{l,n}(0) , h_{l,n}(1) , \ldots , h_{l,n}(N-1) ]^{T}.
\end{align}
Following Jakes' power spectral model of maximum Doppler frequency $f_d$ \cite{Jakes}, the temporal correlation matrix for a time-lag $p$, $\mathbf{R}_{\mathbf{h}_{l}}^{(p)}=\mathbb{E}[ \mathbf{h}_{l,n} \mathbf{h}_{l,n-p}^{H}]$, is given by
\begin{align}
[\mathbf{R}_{h_{l}}^{(p)}]_{k,m}= \sigma_{h_{l}}^{2} J_{0}(2\pi f_{d} T_{s} (k-m+pN_s)).
\end{align} 
 
There exist $N$ samples for each channel tap in every OFDM symbol. This yields a total number of $LN$ samples for the whole channel per OFDM symbol. We use a BEM to reduce the dimensions of the space of parameters required to be estimated. The main goal of the BEM is to accurately model the temporal variation of the channel coefficients within the same OFDM symbol. This variation is approximated by a linear combination of a few bases vectors, $\mathbf{b}_d$, as follows
\begin{align}
\mathbf{h}_{l,n}=\mathbf{B}\mathbf{c}_{l,n}+ \boldsymbol{\nu}_{l,n}
\end{align} 
where $\mathbf{B}=[ \mathbf{b}_{0}, \mathbf{b}_{1}, \ldots , \mathbf{b}_{N_{c}-1} ]$ is the $N \times N_c$ basis matrix containing $N_c$ bases vectors, $\mathbf{c}_{l,n}$ is the $N_c \times 1$ vector of the BEM coefficients corresponding to the $l$th channel tap at the $n$th OFDM symbol and $\boldsymbol{\nu}_{l,n}$ is the modelling error. Thus, the optimal BEM coefficients \cite{Leus} are given by
\begin{align}
&\mathbf{c}_{l,n}=(\mathbf{B}^{H}\mathbf{B})^{-1}\mathbf{B}^{H}\mathbf{h}_{l,n}. \label{BEM_coeff}
\end{align} 
From (\ref{BEM_coeff}), the correlation matrix of the BEM coefficients for a time-lag $p$, $\mathbf{R}_{\mathbf{c_l}}^{(p)}=\mathbb{E}[ \mathbf{c}_{l,n} \mathbf{c}_{l,n-p}^{H}]$ is given by
\begin{align}
\mathbf{R}_{\mathbf{c_l}}^{(p)}=(\mathbf{B}^{H}\mathbf{B})^{-1}\mathbf{B}^{H} \mathbf{R}_{\mathbf{h}_{l}}^{(p)} \mathbf{B} (\mathbf{B}^{H}\mathbf{B})^{-1}.
\end{align} 
In \cite{QR} and \cite{MAP}, (\ref{sysmodel}) is derived in terms of the BEM neglecting the modelling error yielding
\begin{align}\label{measurement_equation}
\mathbf{y}_{n}=\mathbf{S}_{n} \mathbf{c}_{n} + \mathbf{w}_{n}
\end{align}
where the $LN_c \times 1$ vector $\mathbf{c}_{n}$ and the $N \times LN_c$ matrix $\mathbf{S}_{n}$ are given by
\begin{eqnarray}
\mathbf{c}_{n}&=&[ \mathbf{c}_{0,n}^{T}, \mathbf{c}_{1,n}^{T}, \ldots , \mathbf{c}_{L-1,n}^{T} ]^{T} \\
\mathbf{S}_{n}&=&\frac{1}{\sqrt{N}}[ \mathbf{V}_{0,n}, \mathbf{V}_{1,n}, \ldots , \mathbf{V}_{L-1,n} ] \\
\mathbf{V}_{l,n}&=&[\mathbf{D}_{0} \text{diag} \{ \mathcal{X}_{n} \} \mathbf{f}_{l}, \ldots ,\mathbf{D}_{N_{c}-1} \text{diag} \{ \mathcal{X}_{n} \} \mathbf{f}_{l}] 
\end{eqnarray}
$\mathbf{f}_{l}$ is the $l$th column of the DFT matrix $\mathbf{F}$, and the $N \times N$ matrix $\mathbf{D}_{d}$ is given by
\begin{align}
\mathbf{D}_{d}=\text{diag} \{ \mathbf{b}_{d} \} \mathbf{F}^{H}.
\end{align}

\section{Kalman filtering matched to a BEM}
In this section, we introduce the acquisition phase of the proposed algorithm. Then, we briefly present the Kalman filtering approach proposed by \cite{MAP}, where the Kalman filter is matched to a given BEM.
\subsection{Acquisition}
A preamble sequence is transmitted in the beginning of each frame to obtain an initial estimate of the BEM coefficients. Consider $K$ training OFDM symbols transmitted as a preamble. Then, the received preamble signal is given by
\begin{align}
\mathbf{y}=\mathbf{S}\mathbf{c}+\mathbf{w}
\end{align}
where 
\begin{align}
&\mathbf{y}=[\mathbf{y}_{0}^{T}, \ldots , \mathbf{y}_{K-1}^{T}]^{T} \\
&\mathbf{S}= \text{blkdiag} \{ \mathbf{S}_{0}, \ldots , \mathbf{S}_{K-1} \} \\
&\mathbf{c}=[\mathbf{c}_{0}^{T}, \ldots , \mathbf{c}_{K-1}^{T}]^{T} \\
&\mathbf{w}=[\mathbf{w}_{0}^{T}, \ldots , \mathbf{w}_{K-1}^{T}]^{T}.
\end{align}
Thus, the initial LS estimate of the BEM coefficients is
\begin{align}
\hat{\mathbf{c}}=(\mathbf{S}^{H}\mathbf{S})^{-1}\mathbf{S}^{H}\mathbf{y}.
\end{align}

\begin{figure*}[t]
\begin{center}
\includegraphics[width=0.8\textwidth, height=0.43\textwidth]{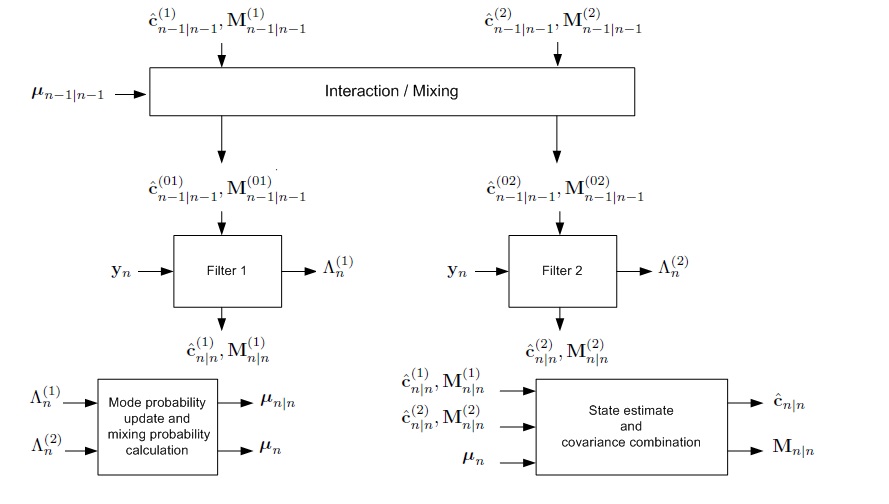}
\caption{Structure of the IMM estimator} \label{Fig1}
\end{center}
\vspace{-5mm}
\end{figure*}

\subsection{Tracking}
Simon et.al proposed a Kalman filter-based tracking for the BEM coefficients in \cite{MAP}. The dynamics of $\mathbf{c}_{l,n}$ is modelled via a first order auto-regressive (AR) model as follows
\begin{align}\label{AR}
\mathbf{c}_{l,n}=\mathbf{A}_{l} \mathbf{c}_{l,n-1} + \mathbf{u}_{l,n}
\end{align}  
where $\mathbf{A}_{l}$ is the transition $N_c \times N_c$ matrix of BEM coefficients of the $l$th channel tap across two consecutive OFDM symbols and $\mathbf{u}_{l,n}$ is the complex Gaussian noise vector of covariance matrix $\mathbf{U}_{l}$. The AR model parameters can be computed through Yule-Walker equations as
\begin{align}
&\mathbf{A}_{l}=\mathbf{R}_{\mathbf{c}_{l}}^{(1)} (\mathbf{R}_{\mathbf{c}_{l}}^{(0)})^{-1} \\
&\mathbf{U}_{l}=\mathbf{R}_{\mathbf{c}_{l}}^{(0)}+ \mathbf{A}_{l}\mathbf{R}_{\mathbf{c}_{l}}^{(-1)}.
\end{align}
Using (\ref{AR}), the AR model for $\mathbf{c}_{n}$ is given by
\begin{align}\label{state_equation}
\mathbf{c}_{n}= \mathbf{A} \mathbf{c}_{n-1} + \mathbf{u}_{n}
\end{align}
where
$\mathbf{A}=\text{blkdiag}\{ \mathbf{A}_0 , \ldots , \mathbf{A}_{L-1} \}$ and the $LN_c \times 1$ vector  
$\mathbf{u}_{n}=[\mathbf{u}_{0,n}^{T} , \ldots , \mathbf{u}_{L-1,n}^{T} ]^{T}$ is the complex Guassian noise vector of covariance matrix $\mathbf{U}=\text{blkdiag}\{ \mathbf{U}_0 , \ldots , \mathbf{U}_{L-1} \}$. Using the state equation (\ref{state_equation}) and the measurement equation (\ref{measurement_equation}), the Kalman filter equations \cite{kay} presented in \cite{MAP} are \\
Time update equations (TUE):
\begin{align}
&\hat{\mathbf{c}}_{n|n-1} = \mathbf{A}\hat{\mathbf{c}}_{n|n-1} \\
&\mathbf{M}_{n|n-1} = \mathbf{A}\mathbf{M}_{n-1|n-1}\mathbf{A}^{H} + \mathbf{U}.
\end{align}
Measurement update equations (MUE):
\begin{align}
&\hat{\mathbf{c}}_{n|n}=\hat{\mathbf{c}}_{n|n-1} + \mathbf{K}_{n} \left( \mathbf{y}_{n}-\mathbf{S}_{n}
\hat{\mathbf{c}}_{n|n-1} \right) \\
&\mathbf{M}_{n|n}=\left[\mathbf{I}_{LN_c}-\mathbf{K}_{n} \mathbf{S}_{n}\right] \mathbf{M}_{n|n-1}. 
\end{align}
where the
filter gain $\mathbf{K}_{n}$ and the innovation covariance matrix $\mathbf{Q}_{n}$ are given respectively by
\begin{eqnarray}
\mathbf{K}_{n}&=&\mathbf{M}_{n|n-1} \mathbf{S}_{n}^{H}\mathbf{Q}_{n}^{-1}\\
\mathbf{Q}_{n}&=&\sigma_{\mathbf{w}}^{2} \mathbf{I}_{N} +
\mathbf{S}_{n} \mathbf{M}_{n|n-1} \mathbf{S}_{n}^{H}.
\end{eqnarray}  
The Kalman filter is implemented in a decision-directed algorithm, where the channel estimate at the ($n-1$)th iteration is used to construct the channel matrix $\mathbf{H}_{n}$ in (\ref{sysmodel}) at the $n$th iteration for equalization. Thus, we get the transmitted symbols $\mathbf{x}_{n}$ which we use to construct the matrix $\mathbf{S}_{n}$.

\section{Interactive Multiple Model Estimator}
In multiple model environments, the optimal state estimate is a function of the elemental state estimates obtained by the estimators tuned to all possible parameter histories. Thus, with time, an exponentially increasing number of filters is required to keep track of all possible model parameter histories. Many suboptimal techniques are proposed to overcome this complexity issue among which the IMM is the most cost efficient.

The structure of the IMM estimator is shown in Fig.~\ref{Fig1}. At time $m$, the state estimate is computed under each possible current model using two filters, with each filter using a different combination of the previous model-conditioned state estimates (mixed initial condition). The model switching process is assumed to be a Markov chain with the known transition probability  matrix
$$ \mathbf{P}=
\begin{bmatrix}
  p^{(11)} & p^{(12)} \\
  p^{(21)} & p^{(22)} 
\end{bmatrix}$$
where $p^{(ij)}$ denotes the transition probability from model $i$ to model $j$.
One cycle of the algorithm consists of the following steps mentioned in\cite{shalom}:
%\vspace{-5mm}
\begin{enumerate}
\item \textbf{Calculation of the mixing probabilities}: The probability that model $i$ is in effect at time ($n-1$) given that model $j$ is in effect at time $n$ and given the measured data up to step ($n-1$) is
\begin{align}
\mu^{(i|j)}_{n-1|n-1}=\frac{1}{\bar{c}^{(j)}}p^{(ij)}\mu^{(i)}_{n-1},~ i,j=1,2
\end{align} 
where $\mu^{(i)}_{n-1}$ is the probability that model $i$ is in effect at time ($n-1$) and the normalization constants are
\begin{align}\label{c_bar}
\bar{c}^{(j)}=\sum_{i=1}^{2}p^{(ij)} \mu^{(i)}_{n-1}.
\end{align}
\item \textbf{Mixing}: Starting with the state estimates of both filters and their associated covariances at time ($n-1$), one computes the mixed initial condition $\hat{\mathbf{c}}^{(0j)}_{n-1|n-1)}$ and $\mathbf{M}^{(0j)}_{n-1|n-1}$ matched to filter $j$ as
\begin{align}
&\hat{\mathbf{c}}^{(0j)}_{n-1|n-1}=\sum_{i=1}^{2} \hat{\mathbf{c}}^{(i)}_{n-1|n-1)} \mu^{(i|j)}_{n-1|n-1} 
\label{state mixed initial condition}\\
&\mathbf{M}^{(0j)}_{n-1|n-1}=\sum_{i=1}^{2}\mu^{(i|j)}_{n-1|n-1} \bigg\{ \mathbf{M}^{(i)}_{n-1|n-1}+ \notag \\
&[\hat{\mathbf{c}}^{(i)}_{n-1|n-1}-\hat{\mathbf{c}}^{(0j)}_{n-1|n-1}]
[\hat{\mathbf{c}}^{(i)}_{n-1|n-1}-\hat{\mathbf{c}}^{(0j)}_{n-1|n-1}]^{H}\bigg\}.
\label{cov mixed initial condition}
\end{align}  
\item \textbf{Mode matched filtering}: The state estimate (\ref{state mixed initial condition})
and its associated covariance (\ref{cov mixed initial condition}) are used as input to the filter matched to model $j$ to yeild $\hat{\mathbf{c}}^{(j)}_{n|n}$ and $\mathbf{M}^{(j)}_{n|n}$.
The likelihood function corresponding to both filters which is the probability of the measurement $\mathbf{y}_{n}$ given that model $j$ is in effect at time $n$ and all the data history up to time ($n-1$) are computed as
\begin{align}
\Lambda^{(j)}_{n}=\frac{1}{|2\pi\mathbf{Q}^{(j)}_{n}|} \exp  \{ & -\frac{1}{2}
[\mathbf{y}_{n}-\hat{\mathbf{y}}^{(j)}_{n|n-1}]^{H} \times \notag  \\     
&( \mathbf{Q}^{(j)}_{n} )^{-1} [ \mathbf{y}_{n}-\hat{\mathbf{y}}^{(j)}_{n|n-1} ] \} 
\end{align}
where $\hat{\mathbf{y}}^{(j)}_{n|n-1}$ is the predicted measurement at time $n$ given data history up to time ($n-1$) and $[\mathbf{y}_{n}-\hat{\mathbf{y}}^{(j)}_{n|n-1}]$ is the innovation at step $n$ with covariance $\mathbf{Q}^{(j)}_{n}$.
\item \textbf{Mode probability update}: Mode probability is updated using the likelihood function as follows
\begin{align}
\mu^{(j)}_{n}=\frac{1}{c} \Lambda^{(j)}_{n} \bar{c}^{(j)}
\end{align}     
where $\bar{c}^{(j)}$ is given by (\ref{c_bar}) and $c$ is a normalization constant given by
\begin{align}
c=\sum_{j=1}^{2}\Lambda^{(j)}_{n}\bar{c}^{(j)}.
\end{align}
\item \textbf{Estimate and covariance combination}: Combination of the model conditioned estimates and their associated covariances is done according to the mixing equations
\begin{equation}
\hat{\mathbf{c}}_{n|n}=\sum_{j=1}^{2}\hat{\mathbf{c}}^{(j)}_{n|n} \mu^{(j)}_{n}
\end{equation}
\begin{equation}
\mathbf{M}_{n|n}\!\!=\!\!\sum_{j=1}^{2}\!\!\mu^{(j)}_{n} \{\mathbf{M}^{(j)}_{n|n}\! + \!  
[\hat{\mathbf{c}}^{(j)}_{n|n}\!-\!\hat{\mathbf{c}}_{n|n}] 
[\hat{\mathbf{c}}^{(j)}_{n|n}\!-\! \hat{\mathbf{c}}_{n|n}]^{H}\}.
\end{equation}

\end{enumerate}

\section{Simulation Results}
In this section, the performance of the proposed algorithm is evaluated in terms of the mean square error (MSE) in the estimate of the most significant channel tap. We consider a SISO-OFDM system employing quadrature phase shift keying (QPSK) with $N=64$, $N_g=N/4$, $N_g=N/16$ and $1/T_s=20$MHz. The transmitted frame consists of $200$ OFDM symbols. The first $K=2$ OFDM symbols of each transmitted frame are used as a preamble sequence utilized to derive the initial channel estimate. Binary Phase shift keying (BPSK) is the modulation technique employed for the preamble. Simulation results are averaged over $10^3$ Monte Carlo runs.

We consider a multipath channel with $4$ taps (equispaced with respect to delay) whose power delay profile is given by $(0,-1,-3,-9)$ dB. Therefore, the maximum delay spread of the channel is given by $4T_s$. The proposed combining technique (IMM) utilizes two complex exponential BEMs with $N_{c}=3$ coefficients each. The first BEM intended for capturing the slow variations in the channel coefficients, within one OFDM symbol, uses a basis matrix
\begin{align}
[\mathbf{B}_L]_{k,m}=e^{j\frac{2\pi}{N}k \left( m-\frac{N_{c}-1}{2} \right)}
\end{align}
while the second BEM intended for capturing the fast variations has a basis matrix
\begin{align}
[\mathbf{B}_H]_{k,m}=e^{j\frac{2\pi}{N}k \left( 2m-(N_{c}-1) \right)}
\end{align}
where the indices $k$ and $m$ start at zero. 

\begin{figure}[t]
\begin{center}
\includegraphics[width=1\columnwidth , height=0.7\columnwidth]{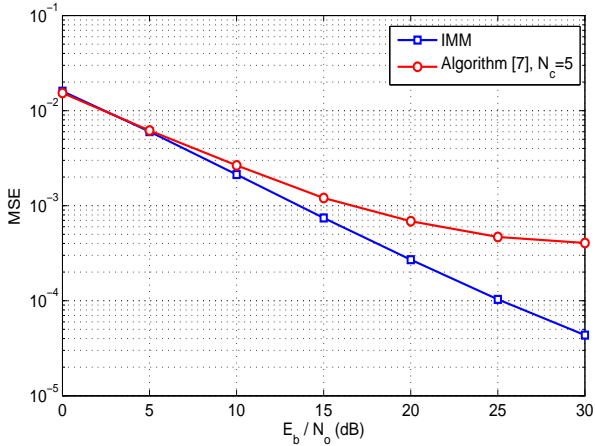}
\caption{MSE versus $E_b/N_0 $} \label{Fig2}
\end{center}
\vspace{-5mm}
\end{figure}

From the definition of the two bases matrices, $\mathbf{B}_L$ and $\mathbf{B}_H$, we note that both are sub-matrices obtained from the N-point DFT matrix $\mathbf{F}$. The $N_c=3$ bases vectors of $\mathbf{B}_L$ correspond to the DC vector and the first two discrete frequencies $\pm \left(\frac{2\pi}{N} \right)$, while the bases vectors of the matrix $\mathbf{B}_{H}$ correspond to the DC vector and the second two discrete frequencies $\pm \left( \frac{4\pi}{N} \right)$. We conjecture that the temporal variation of the channel is better modelled by $\mathbf{B}_L$ in low Doppler scenarios, while $\mathbf{B}_{H}$ performs better in high Doppler scenarios. Thus, the proposed IMM-based combining technique is expected to perform well in a wide span of possible communication scenarios. The performance of the IMM is compared to algorithm \cite{MAP} utilizing a single Kalman filter matched to a basis matrix with $N_c=5$ composed of all vectors in $\mathbf{B}_{L}$ and $\mathbf{B}_{H}$. The IMM is initialized with $\mathbf{P}=\frac{1}{2}\mathbf{I}_{2}$ and $\mu^{(1)}_{0},\mu^{(2)}_{0}=\frac{1}{2}$ to indicate no prior knowledge of the channel.

For illustration purposes, the random multipath channel is generated according to the BEMs. The channel follows $\mathbf{B}_{L}$ in the first half of the transmitted frame (first $100$ OFDM symbols), while it follows $\mathbf{B}_{H}$ in the second half of the frame (second $100$ OFDM symbols). The MSE of the estimate of the $l$th channel tap is defined as
\begin{align}
\text{MSE}_{l}=\frac{\parallel \mathbf{h}_{l}-\hat{\mathbf{h}}_{l} \parallel^{2}}{200N}
\end{align} 
where $\mathbf{h}_{l}$ denotes the $200N \times 1$ vector of the $l$th channel tap.

The MSE in the estimate of the most significant tap is plotted in Fig.~\ref{Fig2} versus $E_b/N_0$. It is shown that the proposed IMM outperforms algorithm \cite{MAP} with $N_c=5$ (with bases vectors corresponding to all frequencies available in $\mathbf{B}_{L}$ and $\mathbf{B}_{H}$). This can be explained as follows: The number of parameters to be estimated using the IMM is $LN_c=12$, which is less than $LN_c=20$ in the case of algorithm \cite{MAP} with larger basis given the same $N$-dimensional measurement set $\mathbf{y}_{n}$. Thus, the Cramer Rao bound is in favour of the IMM.	

\begin{figure}[t]
\begin{center}
\includegraphics[width=1\columnwidth , height=0.7\columnwidth]{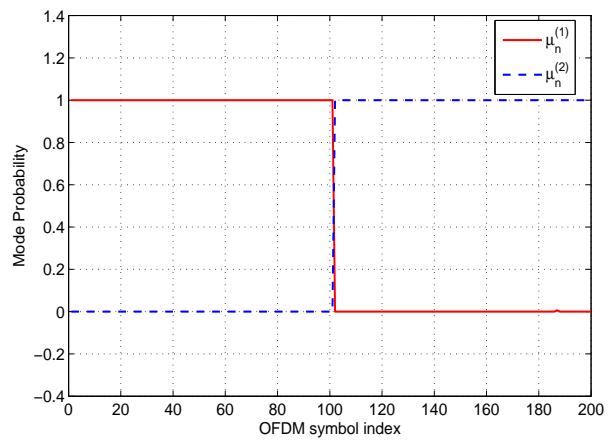}
\caption{Mode probability versus OFDM symbol index} \label{Fig3}
\end{center}
\vspace{-5mm}
\end{figure}

The dynamics of the proposed IMM is shown for one transmitted frame in Fig.~\ref{Fig3}. The two mode probabilities $\{\mu^{(j)}_{n}\}_{j=1}^{2}$ are plotted versus the OFDM symbol index. Fig.~\ref{Fig3} shows the capability of the IMM to lock to the true BEM. The sharp transition in the middle of the frame illustrates the enhanced sensitivity of the IMM to variations of the channel dynamics within the same frame. 

\section{Conclusion}
In this paper, we propose an algorithm for channel estimation, acquisition and tracking, using parametric channel models. The temporal variation of the channel coefficients within one OFDM symbol is modelled by two BEMs. Kalman filtering is employed to track the coefficients of both BEMs. Then, an IMM estimator is used to mix the estimates obtained via each Kalman filter. The message behind the paper is to show that we can achieve higher estimation accuracy through dynamically mixing the estimates obtained via different BEMs. These BEMs are not limited to $\mathbf{B}_{L}$ and $\mathbf{B}_{H}$. The same approach generalizes to any two (or more) basis matrices. Moreover, the algorithm adapts itself to dramatic variations in the underlying environment even within the same OFDM symbol. 
Numerical simulations signify the gain obtained through using the proposed combining technique. 

\bibliographystyle
{IEEEtran}
\bibliography{IEEEabrv,ref}

\begin{thebibliography}{10}
\providecommand{\url}[1]{#1}
\csname url@samestyle\endcsname
\providecommand{\newblock}{\relax}
\providecommand{\bibinfo}[2]{#2}
\providecommand{\BIBentrySTDinterwordspacing}{\spaceskip=0pt\relax}
\providecommand{\BIBentryALTinterwordstretchfactor}{4}
\providecommand{\BIBentryALTinterwordspacing}{\spaceskip=\fontdimen2\font plus
\BIBentryALTinterwordstretchfactor\fontdimen3\font minus
  \fontdimen4\font\relax}
\providecommand{\BIBforeignlanguage}[2]{{%
\expandafter\ifx\csname l@#1\endcsname\relax
\typeout{** WARNING: IEEEtran.bst: No hyphenation pattern has been}%
\typeout{** loaded for the language `#1'. Using the pattern for}%
\typeout{** the default language instead.}%
\else
\language=\csname l@#1\endcsname
\fi
#2}}
\providecommand{\BIBdecl}{\relax}
\BIBdecl

\bibitem{RLS}
H.~Nguyen-Le, T.~Le-Ngoc, and C.~Ko, ``{RLS}-based joint estimation and
  tracking of channel response, sampling, and carrier frequency offsets for
  {OFDM},'' \emph{IEEE Transactions on Broadcasting}, vol.~55, no.~1, pp.
  84--94, December 2009.

\bibitem{Nafouri}
N.~A.-D. Tareq Y Al-Naffouri, KM Zahidul~Islam and S.~Lu, ``A model reduction
  approach for {OFDM} channel estimation under high mobility conditions,''
  \emph{IEEE Transactions on Signal Processing}, vol.~58, no.~4, pp.
  2181--2193, 2010.

\bibitem{Giannakis}
X.~Ma, G.~B. Giannakis, and S.~Ohno, ``Optimal training for block transmissions
  over doubly selective wireless fading channels,'' \emph{IEEE Transactions on
  Signal Processing}, vol.~51, no.~5, pp. 1351--1366, 2003.

\bibitem{subblock}
S.~He and J.~K. Tugnait, ``Doubly-selective channel estimation using
  exponential basis models and subblock tracking,'' in \emph{Global
  Telecommunications Conference, 2007. GLOBECOM'07. IEEE}.\hskip 1em plus 0.5em
  minus 0.4em\relax IEEE, 2007, pp. 2847--2851.

\bibitem{QR}
H.~Hijazi and L.~Ros, ``Joint data {QR}-detection and {K}alman estimation for
  {OFDM} time-varying {R}ayleigh channel complex gains,'' \emph{IEEE
  Transactions on Communications}, vol.~58, no.~1, pp. 170--178, 2010.

\bibitem{KL_BEM}
M.~Visintin, ``Karhunen-{L}oeve expansion of a fast {R}ayleigh fading
  process,'' \emph{Electronics Letters}, vol.~32, no.~18, p. 1712, 1996.

\bibitem{MAP}
E.~P. Simon, M.~Berbineau, and M.~Li{\'e}nard, ``Joint {CFO} and channel
  acquisition and tracking based on parametric channel modeling for {OFDM}
  systems in the presence of high mobility,'' in \emph{2011 11th International
  Conference on Telecommunications (ITST)}.\hskip 1em plus 0.5em minus
  0.4em\relax IEEE, 2011, pp. 565--570.

\bibitem{Jakes}
W.~C. Jakes, \emph{Microwave {M}obile {C}ommunications}.\hskip 1em plus 0.5em
  minus 0.4em\relax IEEE press, 1983.

\bibitem{Leus}
G.~Leus, ``On the estimation of rapidly time-varying channels,'' in \emph{Proc.
  EUSIPCO}, 2004, pp. 2227--2230.

\bibitem{kay}
S.~Kay, \emph{Fundamentals of {S}tatistical {S}ignal {P}rocessing, {V}olume
  {I}: {E}stimation {T}heory}.\hskip 1em plus 0.5em minus 0.4em\relax Englewood
  Cliffs, NJ: Prentice Hall, 1993.

\bibitem{shalom}
Y.~Bar-Shalom, X.~Li, and T.~Kirubarajan, \emph{Estimation with applications to
  tracking and navigation: theory algorithms and software}.\hskip 1em plus
  0.5em minus 0.4em\relax Wiley-Interscience, 2001.

\end{thebibliography}
\end{document}